# Thermo-quantum diffusion in periodic potentials

Roumen Tsekov
Department of Physical Chemistry, University of Sofia, 1164 Sofia, Bulgaria

Quantum Brownian motion in a periodic cosine potential is studied and a simple estimate of the tunneling effect is obtained in the frames of a semiclassical quasi-equilibrium approach. It is shown that the latter is applicable for heavy particles but electrons cannot be described properly since the quantum effects dominate over the thermal ones. The electron purely quantum diffusion is investigated at zero temperature and demonstrates that electrons do not obey the classical Einstein law of Brownian motion in the field of periodic potentials, since the dispersion of the Gaussian wave packet increases logarithmically in time.

Diffusion in the field of periodic potentials is important for particle transport via ordered structures, typical in condensed matter. The underlying mechanism is Brownian motion. If the diffusing particles are quantum ones, e.g. electrons, protons, light atoms, etc., their migration is described via the theory of quantum Brownian motion [1, 2], which is significantly affected by the tunneling effect [3]. Hereafter an alternative approach to the problem of quantum Brownian motion in periodic potentials is developed based on the Bohm quantum potential [4]. In contrast to the WKB theory valid in vacuum the present paper describes tunneling in a dissipative environment. The starting point is a nonlinear quantum Smoluchowski-like equation [5, 6]

$$\partial_t \rho = \partial_x [\rho \partial_x \int_0^\beta (V+Q)_b d\beta + \partial_x \rho]/\beta b \qquad (1)$$

describing the overdamped evolution of the probability density $\rho(x,t)$ to find a quantum particle at a given place under the action of an arbitrary external potential $V$. Here $b$ is the particle friction coefficient, $\beta = 1/k_B T$ and $Q \equiv -\hbar^2 \partial_x^2 \sqrt{\rho} / 2m\sqrt{\rho}$ is the Bohm quantum potential with $m$ being the particle mass. Thus, due to the quantum potential the diffusion becomes mass-sensitive. The goal of the present analysis is to estimate the diffusion coefficient from Eq. (1) for the case of a periodic cosine potential $V(x) = U\cos(qx)$.

Since Eq. (1) is nonlinear due to the quantum potential, obtaining its general solution is mathematically frustrated. In a semiclassical approach one could estimate $Q$ by using the classical solution of Eq. (1). However, even in the classical case it is impossible to obtain a simple physically transparent expression for the probability density. For this reason, we are going to apply a stronger approximation by employing the equilibrium classical Boltzmann distribution $\rho_B \sim \exp(-\beta V)$ in the quantum potential to obtain

$$Q(\rho_B) = (\beta\hbar^2/4m)[\partial_x^2 V - \beta(\partial_x V)^2/2] = -\lambda_T^2 q^2 [V + \beta(U^2 - V^2)/2] \qquad (2)$$

The first expression here is general, while the last one is valid for the present particular cosine potential; the thermal de Broglie wave length is introduced by the expression $\lambda_T \equiv \hbar/2\sqrt{mk_B T}$. Note that because of this quasi-equilibrium approximation the quantum potential (2) becomes temperature-dependent. Substituting Eq. (2) in Eq. (1) and performing the integration over $\beta$ results in the following semiclassical quasi-equilibrium Smoluchowski equation

$$\partial_t \rho = \partial_x (\beta \rho \partial_x W + \partial_x \rho)/\beta b \qquad (3)$$

where the new effective potential is given by $W \equiv [1 - \lambda_T^2 q^2 (1 - \beta V/3)/2]V$ and possesses the same periodicity as the external potential $V$. Owing to the Boltzmann distribution employed, Eq. (3) is a quasi-equilibrium one and in the case of a free quantum Brownian particle ($V = 0$) it reduces to the classical diffusion equation. Hence, its application is restricted to relatively strong external potentials, where the semiclassical quantum effects still play a role. For instance, in a harmonic potential $V = m\omega_0^2 x^2/2$ the effective potential $W = [1 - (\beta\hbar\omega_0/2)^2/3]V$ is harmonic as well. Thus, the average equilibrium position dispersion, following from Eq. (3), $\sigma_x^2 = 1/\beta m\omega_0^2 [1 - (\beta\hbar\omega_0/2)^2/3]$ is the high temperature approximation of the exact thermodynamic expression $\sigma_x^2 = (\hbar/2m\omega_0)\coth(\beta\hbar\omega_0/2)$.

Lifson and Jackson [7] and later Festa and d'Agliano [8] have derived a general formula for calculation of the diffusion coefficient $D$ from Eq. (3) applied for periodic potentials

$$1/D = \beta b <\exp(\beta W)><\exp(-\beta W)> \qquad (4)$$

where the brackets $<\cdot>$ indicate spatial geometric average. At room temperature one can usually neglect the small nonlinear contribution of the external potential $V$ to $W$ and thus the latter acquires the simplified form $W = (1 - \lambda_T^2 q^2/2)V$. In this case the spatial averaging in Eq. (4) can be analytically accomplished and the result reads

$$1/D = \beta b I_0^2 [\beta(1 - \lambda_T^2 q^2/2)U] \qquad (5)$$

where $I_0$ is the modified Bessel function of first kind and zero order. In the case of free Brownian motion ($U = 0$) Eq. (5) provides the classical Einstein formula $D = 1/\beta b$, which is due to the semiclassical quasi-equilibrium approach [9]. At large arguments the modified Bessel function can be approximated well as $I_0(x) \approx \exp(x)/\sqrt{2\pi x}$ and Eq. (5) reduces in this case to the Arrhenius law

$$D = \pi(2 - \lambda_T^2 q^2)(U/b)\exp[-\beta(2 - \lambda_T^2 q^2)U] \tag{6}$$

As is seen, both the effective activation energy $E_a \equiv (2 - \lambda_T^2 q^2)U$ and the pre-exponential factor $\pi(2 - \lambda_T^2 q^2)U/b$ from Eq. (6) are affected by the quantum effect. In the classical limit $\lambda_T = 0$ and the activation energy $E_a = 2U$ equals to the difference between the maximal and minimal value of the external potential. The quantum effect decreases effectively the activation energy due to the quantum tunneling. For instance, the thermal de Broglie wave length for a proton at room temperature equals to $\lambda_T \approx 0.2$ Å. If protons are diffusing in a structured medium with a lattice constant 3 Å then $\lambda_T^2 q^2 / 2 \approx 0.1$. Hence, the tunneling effect will decrease the activation energy by 10 % as well as the pre-exponential factor. Note that if $\lambda_T^2 q^2 = 2$ the diffusion is free since $W = 0$, which corresponds for a proton to $T \approx 25$ K. For electrons the thermal de Broglie wave length at room temperature is $\lambda_T \approx 8.6$ Å and hence $\lambda_T^2 q^2 / 2 \approx 162$. Since this number is much larger than 1, the conclusion is that the theory above is not applicable to electrons since they are essentially quantum particles and cannot be described by a quasi-equilibrium semiclassical approach, unless they are moving in very flat potentials with $q < 1/\lambda_T$.

For electrons one should solve directly Eq. (1) instead of Eq. (3). To simplify now the problem we will consider the case of zero temperature, where Eq. (1) reduces to

$$\partial_t \rho = \partial_x[\rho \partial_x(V + Q)]/b = \partial_x[\rho \partial_x V / b - \hbar^2 \partial_x(\rho \partial_x^2 \ln \rho)/4mb] \tag{7}$$

This is still a nonlinear equation, which is not easy to solve. To linearize it one can use the fact that the effective shape of the probability density is expected to be Gaussian in the case of relatively low $U$, which corresponds to lack of localization. Hence, one can employ the approximation $\partial_x^2 \ln \rho = -1/\sigma_x^2$, where $\sigma_x^2$ is the dispersion of the probability density. Introducing this expression in Eq. (7) yields a Smoluchowski equation

$$\partial_t \rho = \partial_x(\rho \partial_x V + \beta_Q^{-1} \partial_x \rho)/b \tag{8}$$

with a new quantum temperature $\beta_Q^{-1} \equiv \hbar^2/4m\sigma_x^2$ corresponds to the momentum dispersion from the minimal Heisenberg relation valid for Gaussian processes. Since $\beta_Q$ depends on time via $\sigma_x^2$ one should use a more general expression of the Festa-d'Agliano formula [8]

$$\partial_t \sigma_x^2 / 2 = [\beta_Q b < \exp(\beta_Q V) >< \exp(-\beta_Q V) >]^{-1} = 1/\beta_Q b I_0^2(\beta_Q U) \qquad (9)$$

Integration of Eq. (9) on time yields the relation

$$(\beta_Q U)^2 [I_0^2(\beta_Q U) - I_1^2(\beta_Q U)] = 16 m U^2 t / \hbar^2 b \qquad (10)$$

where $I_1$ is the modified Bessel function of first kind and first order.

If $U=0$, Eq. (10) reduces to the already known expression $\sigma_x^2 = \hbar\sqrt{t/mb}$ for the purely quantum diffusion in a non-structured environment [6]. Note that this dispersion does not depend linearly on time and hence no diffusion constant exists. In the opposite case $2\beta_Q U > 1$ the modified Bessel functions difference is well approximated by $I_0^2(x) - I_1^2(x) \approx \exp(2x)/2\pi x^2$. Thus according to Eq. (10) the dispersion of the purely quantum diffusion in a strong cosine potential depends logarithmically on time

$$\sigma_x^2 = (\hbar^2/8mU) \ln(32\pi m U^2 t / \hbar^2 b) \qquad (11)$$

This result demonstrates again the lack of classical diffusive behavior. As seen from Eq. (11) the magnitude of the deviation $\sigma_x$ scales with the de Broglie wave length $\lambda_U = \hbar/2\sqrt{2mU}$ of the activation energy [9], while its relaxation time $b/m\omega_U^2$ corresponds to that of a harmonic oscillator with an own frequency $\omega_U = 4U/\hbar$. Even if this logarithmic dependence is derived here particularly for a cosine external potential we believe that it is general for any periodic potential and reflects the Arrhenius law. Indeed the latter holds in any periodic potential when the potential barriers are much higher than the thermal energy and the particular type of the potential affects the pre-exponential factor only [7, 8]. Hence, the quantum diffusion does not obey the classical linear Einstein law of Brownian motion in periodic potentials as well [10]. The account for the nonlinear term in the potential $W$ will introduce some aperiodic behavior, which certainly will accelerate the diffusion process [11].

In the case of a free quantum Brownian particle ($V=0$) an alternative way to simplify Eq. (7) is, instead of the quantum pressure, to linearize the quantum potential around the homogeneous equilibrium distribution. In this way, Eq. (7) reduces to

$$\partial_t \rho = -\hbar^2 \partial_x^4 \rho / 4mb \tag{12}$$

This equation is not parabolic one and hence it does not describe an ordinary diffusion process. Since the Fourier image $\rho_q = \exp(-\hbar^2 q^4 t / 4mb)$ of the solution of Eq. (12) corresponds to a probability density, which is not positively defined, it seems that Eq. (12) is a worse approximation than the diffusive Eq. (8). The solution of $\partial_t \rho = \hbar^2 \partial_x^2 \rho / 4mb\sigma_x^2$ is a Gaussian distribution density with dispersion $\sigma_x^2 = \hbar\sqrt{t/mb}$. Equation (7) is also exactly solvable for a free quantum Brownian particle and its solution is, of course, the same Gaussian distribution [6].